\documentclass[aps,showpacs,floats,amsmath,amssymb]{revtex4}
\usepackage{graphicx}
\usepackage[next]{inputenc}
\usepackage{epsfig}
\def\be{\begin{equation}}
\def\ee{\end{equation}}

\def\bi{\begin{itemize}}
\def\ei{\end{itemize}}
\def\bn{\begin{enumerate}}
\def\en{\end{enumerate}}
\def\bea{\begin{eqnarray}}
\def\eea{\end{eqnarray}}
\def\no{\nonumber}
\def\ba{\begin{array}}
\def\ea{\end{array}}
\def\bd{\begin{displaymath}}
\def\ed{\end{displaymath}}
\def\la{\langle}
\def\ra{\rangle}
\def\gap{E_g}
\def\gapx{E^x_g}
\def\gapy{E^y_g}
\def\gapz{E^z_g}

\def\ket#1{{\vert #1 \rangle}}

\begin{document}
\title{Antiferromagnetic and spin gap phases of the anisotropic Kondo necklace model}

\author{A. Langari$^{1,2,3}$ and P. Thalmeier$^2$}
\affiliation{$^1$Physics Department, Sharif University of Technology, Tehran 11365-9161, Iran\\
$^2$Max-Planck-Institut f\"ur Chemische Physik fester Stoffe, 01187 Dresden, Germany\\
$^3$Institute for Studies in Theoretical Physics and Mathematics (IPM), Tehran 19395-5531, Iran 
}
\date{\today}
\begin{abstract}
We have studied the effect of anisotropies on the quantum phase
transition of the Kondo necklace model in dimensions D=1, 2 and 3. Both the
anisotropy $\delta$ of the inter-site interaction term and anisotropy
$\Delta$ of the on-site Kondo interaction have been included. We use a
bond operator method with constraints implemented in mean field
approximation. Starting from the paramagnetic phase we determine the
critical ratio $(t/J)_c$ of the quantum critical point and associated
scaling exponents of the Kondo-singlet gap. We show that in the case
of easy-axis type anisotropy $\delta >1$ a qualitatively new
behavior in comparison to the conventional Kondo necklace model with
($\delta$,$\Delta$)=(0,1) appears. We have also obtained the antiferromagnetic
order parameter in the long range ordered phase for $t > t_c$.
\leftskip 2cm \rightskip 2cm
\end{abstract}
\maketitle

\section{Introduction}
\label{sec1}

Quantum phase transitions in heavy fermion systems
between a Fermi liquid and antiferromagnetically
(AF) ordered state have been at the focus of research recently 
\cite{Sachdevbook,Vojta03}. This is due to
mainly two reasons: Firstly in the vicinity of the quantum critical
point (QCP) which is usually reached by applying external or
doping-induced chemical pressure, the intrinsic energy scale of the
compound, i.e., effective Fermi temperature T$^*$ or N\'eel temperature T$_N$
vanishes. This leads to a breakdown of the Fermi liquid picture and
non-Fermi liquid behavior of thermodynamic
and transport quantities is observed close to the
QCP \cite{Continentinobook}. 
It is therefore of great interest to understand the approach to
the quantum critical region within suitable theoretical models. The
most important among them is the Kondo lattice model consisting of a
free conduction band and an on-site AF Kondo interaction which favors
nonmagnetic singlet formation. In  the second order it also leads to
the effective inter-site RKKY interactions which favor magnetic
order. Their competition leads to the appearance of the QCP. In such a
picture only spin degrees of freedom are involved in
the quantum phase transition. Therefore the Kondo lattice model may be
replaced by a simpler model where the itinerant hopping part is
simulated by an inter-site interaction of the itinerant spins. In the
work of Doniach \cite{Doniach77} it was shown that in dimension D=1
this replacement, leading to the 'Kondo-necklace' model is indeed
exact assuming that the inter-site interaction is of the
XY-type. Later more general anisotropic inter-site interactions have
been considered including the z-component of conduction-electron spins
\cite{Yamamoto02}. In the itinerant picture this would be equivalent
to adding an interaction term for conduction electrons,
i.e., considering Kondo spins that are screened by correlated
conduction electrons \cite{Schork99}. In addition the Kondo-necklace model
was considered in higher dimension \cite{Zhang00}. In this case the
direct connection to the original
itinerant Kondo lattice model is lost. One then has to consider the
Kondo necklace model for D $\geq$ 2 as a model in its own right. In
fact exact diagonalisation results \cite{Zerec05} for finite clusters
show that the two models are still closely related even for D = 2. 

In the present work we want to study the possible quantum phase
transition in the D = 1-3 Kondo-necklace type model under rather
general assumption of the anisotropy of both the inter-site
interaction and on-site Kondo terms. They are characterised by the pair of
parameters ($\delta,\Delta$). For the former $\delta \neq$ 0
describes the deviation from the XY- type interaction associated with
free conduction electrons in 1D and for the latter $\Delta \neq$ 1
describes deviation from the isotropic on-site Kondo interaction. The
$\Delta$ anisotropy is always present in real Kondo compounds like
Ce-based intermetallics due to the crystalline electric field (CEF).

For uniaxial symmetry it splits the J $=\frac{5}{2}$ multiplet into a
sequence of Kramers doublets. The ground state doublet $|\pm\ra$ may
be described by a pseudo-spin S=$\frac{1}{2}$. Projecting the Kondo
term to this subspace then leads to different on-site exchange for
S$_{x,y}$ and S$_z$ components. This defines a local anisotropy ratio
$\Delta$ = J$_z$/J$_x$ of the pseudo-spin Kondo term. In real Kondo
compounds this ratio may assume any value between the easy axis
(Ising-like) case $\Delta\rightarrow\infty$ and the easy-plane
(xy-like) case $\Delta$ = 0. Examples of tetragonal compounds are
CeRu$_2$Si$_2$ \cite{Saha02}
for the former and YbRh$_2$Si$_2$ \cite{Trovarelli00} for the latter.

We study the quantum phase transition from the paramagnetic (Kondo-singlet)
side as function of the control paramter $t/J$ giving the ratio of
inter-site to on-site interaction strength and as function of the
ansisotropies ($\delta$,$\Delta$). Previously this has only been
performed in the special case  ($\delta$,$\Delta$)=(0,1)
\cite{Zhang00}. For the case of general ($\delta$,$\Delta$) we use the
same bond operator representations of spin variables where on-site
constraints are implemented in mean field approximation.

In Sec.~\ref{sec2} we give a brief definition of the anisotropic Kondo
necklace model and the bond-operator transformation. In Sec.~\ref{sec3} 
the ground state energy and excitation spectrum is calculated in mean
field approximation. Using the numerical results the quantum phase transitions
and phase diagrams are discussed in Sec.~\ref{sec4}. The extension of the mean field
approach for the antiferromagnetic ordered phase is presented in Sec.~\ref{af-phase}.
Finally Sec.~\ref{sec5} gives the summary.


\section{Anisotropic Kondo necklace model 
\label{sec2}}

We investigate the anisotropic Kondo necklace model which is defined
by the following Hamiltonian
\be
H=H_t+H_J=t\sum_{<i, j>} (\tau_i^x \tau_j^x + 
\tau_i^y \tau_j^y + \delta \tau_i^z \tau_j^z)
+J\sum_{i}(\tau_i^x S_i^x+\tau_i^y S_i^y + 
\Delta \tau_i^z S_i^z) .
\label{hamiltonian}
\ee
In the above Hamiltonian, $\tau_n^{\alpha}$ represent the
$\alpha$-component of spin of the itinerant electron at site $n$ and
$S_n^{\alpha}$ is the $\alpha$-component of localized spins at
position $n$. For the exchange coupling between the itinerant and
localized spins we have used $J_x\equiv J=1$ as the reference
energy scale in all figures. The hopping
parameter of the itinerant electrons is proportional to $t$ with the
anisotropy in the z-direction given by $\delta$. Therefore the above
model has three control parameters: $t/J$ and the anisotropies
($\delta$,$\Delta$). 
%
\begin{figure}[tbc]
\begin{center}
\includegraphics[width=9cm,angle=-90]{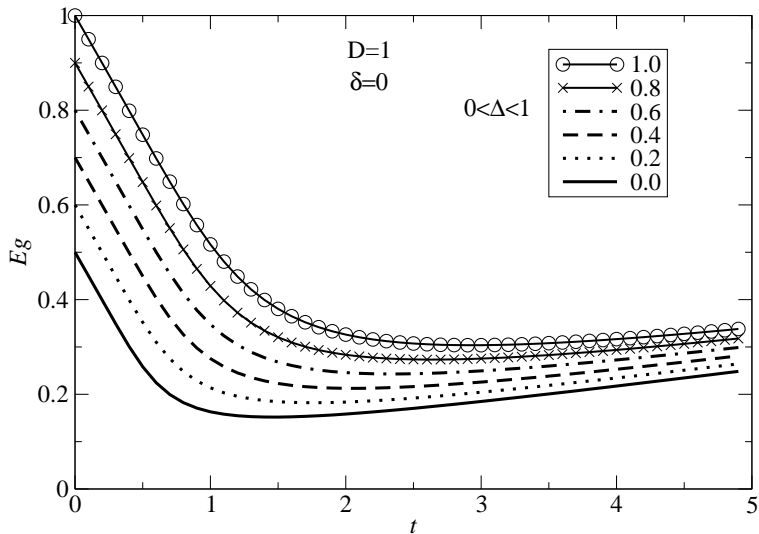}
\caption{The energy gap $\gap$ versus hopping strength in the one dimensional
lattice. 
Different plots show the effect of local anisotropy, $0 \leq \Delta \leq 1$.}
\label{fig1}
\end{center}
\end{figure}
%
There are two main reasons for considering the anisotropy both in the
itinerant and localized parts of the interactions: ({\it i}) In real
materials where the Kondo effect has been observed there exists the
anisotropy in the interaction between the localized and itinerant
electrons.  ({\it ii}) The anisotropy in the itinerant part of the
interaction enables us to investigate the effect of symmetry on the
results of the Kondo necklace model compared with the Kondo model. In
the present model the case of $\delta\neq 1$ has $U(1)$ symmetry and
for $\delta = 1$ it has SU(2) symmetry, the same as in the genuine Kondo model. 
Moreover, in one dimensional lattice nonzero $\delta$ represents
the interaction between itinerant electrons \cite{Spronken81}.

Our approach is based on the strong coupling limit $J/t \rightarrow \infty$($t=0$) where the
model is composed of independent pairs of spins ($ \tau,\mathbf{S}$). The
Hilbert space of a pair of spins consists of 4 states where it can be
represented by a singlet and a triplet. This basis can be created out
of the vacuum by singlet and triplet creation operators:
$\ket{s}=s^{\dagger}\ket{0}$ and $\ket{t}=t_{\alpha}^{\dagger}\ket{0}$
($\alpha=x, y, z$). In terms of singlet-triplet operators the spin
operator of the localized and conduction electrons are given by
\bea
S_{n, \alpha}=\frac{1}{2}(s^{\dagger}_n t_{n, \alpha}+
t_{n, \alpha}^{\dagger} s_n -i \epsilon_{\alpha \beta \gamma} 
t_{n, \beta}^{\dagger}t_{n, \gamma}), \nonumber \\
\tau_{n, \alpha}=\frac{1}{2}(-s^{\dagger}_n t_{n, \alpha}-
t_{n, \alpha}^{\dagger} s_n -i \epsilon_{\alpha \beta \gamma} 
t_{n, \beta}^{\dagger}t_{n, \gamma}),
\label{s-tau}
\eea
where ($\alpha, \beta, \gamma$) represent the ($x, y, z$) components
and $\epsilon$ is the totally antisymmetric tensor. This type of
representation has been introduced first by Sachdev and Bhatt
\cite{Sachdev90} called {\it bond operators}.  The bond operators
satisfy bosonic commutation relations $[s_n, s_n^{\dagger}]=1$,
$[t_{n, \alpha}, t_{n, \beta}^{\dagger}]=\delta_{\alpha, \beta}$ and
$[s_n, t_{n, \beta}^{\dagger}]=0$. The physical states are obtained by
the local constraint $s^{\dagger}_n s_n+\sum_{\alpha} t_{n, \alpha}^{\dagger} t_{n, \alpha}=1$.
Using Eq.(\ref{s-tau}) the Hamiltonian (Eq.(\ref{hamiltonian})) may be
expressed in terms of bond operators. As discussed in the next section
it then reduces to a simple form within the mean-field approximation.
%
\begin{figure}[tbc]
\begin{center}
\includegraphics[width=9cm,angle=-90]{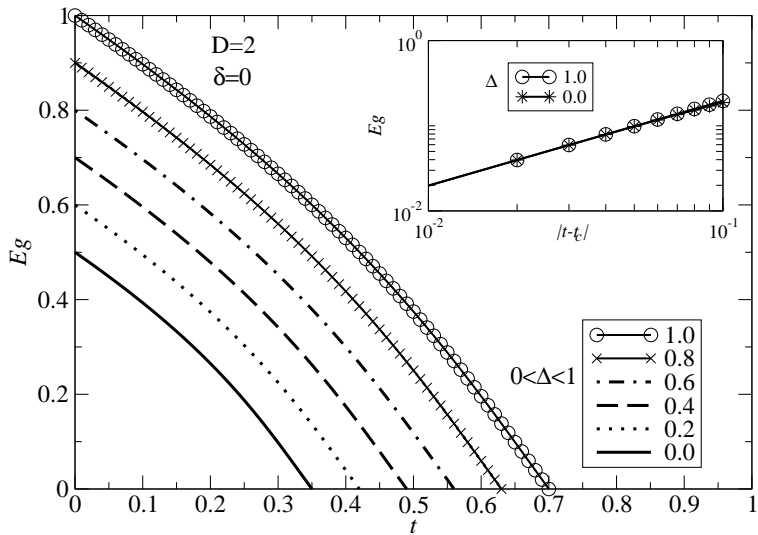}
\caption{The energy gap $\gap$ versus hopping strength in the two-dimensional
lattice for $\delta=0$. 
The inset shows the log-log plot of $E_g$ versus $|t-t_c|$ for $\Delta=0, 1$ where
the gap scales like  $\gap\sim |t-t_c|^\nu$ with $\nu\simeq 1$ close to the critical 
point.}
\label{fig2}
\end{center}
\end{figure}
%
\section{Singlet condensation and ground state properties in
mean-field approximation}
\label{sec3}

We have followed the mean-field approach introduced in
Ref.\onlinecite{Zhang00} where the local constraint is
replaced by a global one for the average amplitudes $\la s_n\ra$ and  
$\la t_{n, \alpha} \ra$.  In this respect we start from the strong coupling
limit $ J/t \rightarrow \infty$ ($t=0$) where the ground state is composed of the direct product
of local singlets. In this limit we have $\la s_n \ra=\la
s_n^{\dagger} \ra=1$ and $\la t_{n, \alpha} \ra=\la t_{n,
\alpha}^{\dagger} \ra=0$ where the ground state represents the pure
condensation of local singlets. Turning on the hopping of conduction
electrons ($t \neq 0$) the triplet occupation at each site becomes
nonzero but still very small as will be shown later. Thus, for a small
value of $t/J$  we consider a mean field
value for $\la s_n \ra=\la s_n^{\dagger} \ra=\bar{s}$ and the
excitations above this background are defined by the triplet states
which define the mean field Hamiltonian.
%
\begin{figure}[tbc]
\begin{center}
\includegraphics[width=9cm,angle=-90]{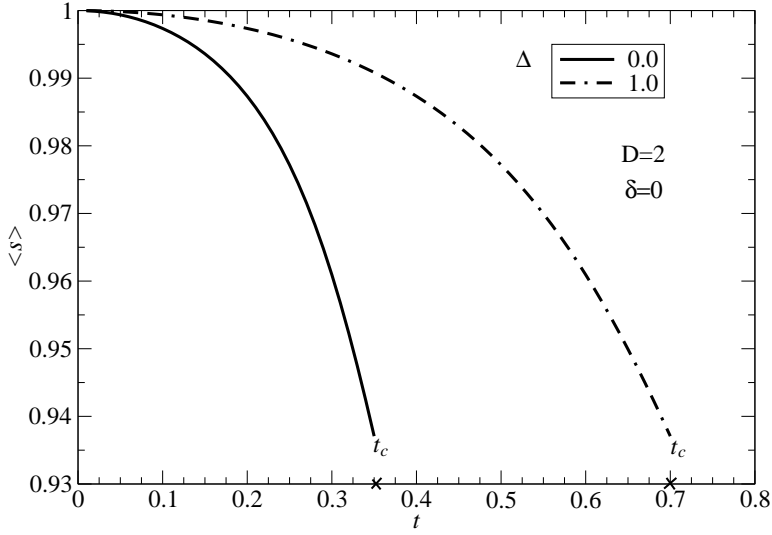}
\caption{The occupation of singlets in the ground state ($<s>=\bar{s}$) versus the 
hopping strength up to the critical point $t_c$ for two different
values of anisotropy $\Delta=0, 1$ in the two dimensional lattice.  The crosses
($\times$) indicate the critical point for each case. 
The deviation from fully occupied state ($<s>=1$) is very small for $t < t_c$.}
\label{fig3}
\end{center}
\end{figure}
%
The hopping term of the Hamiltonian which is
modeled by the XXZ interaction between the spin of conduction
electrons will be written as a sum of three terms, i.e., $H_t= H_1 + H_2 + H_3$
associated with the different types of singlet and triplet
interactions. The first term is given by
\begin{eqnarray}
H_1=\frac{t}{4}\sum_{<n, m>}\sum_{ \alpha=x, y} \Big(s_n^{\dagger} t_{n, \alpha}
(s_m^{\dagger} t_{m, \alpha}+t_{m, \alpha}^{\dagger}s_m)+ h.c.\Big)\\
+\frac{t \delta}{4}\sum_{<n, m>}\Big(s_n^{\dagger} t_{n, z}
(s_m^{\dagger} t_{m, z}+t_{m, z}^{\dagger}s_m)+ h.c.\Big).
\end{eqnarray}
The explicit form of $H_2$ and $H_3$ are given in the
appendix. However, $H_2$ contains the interaction between the triplet
bosons which will have negligible effect on the result of  considering only $H_1$
because the average triplet occupation is small. The contribution of
$H_3$ to the ground state energy is identically zero in mean field
approximation, since each term of $H_3$ consists of three triplets and
one singlet operator.

The exchange term ($H_J$) is diagonal in terms of the bond operators,
\begin{eqnarray}
H_J=\frac{J}{4}\sum_n \Big( -(2+\Delta)s_n^{\dagger}s_n +
(2-\Delta)t_{n,z}^{\dagger}t_{n,z}
+\Delta(t_{n,x}^{\dagger}t_{n,x}+t_{n,y}^{\dagger}t_{n,y})\Big).
\end{eqnarray}
%
The physical
constraint is imposed by adding a Lagrange term at each site with an
associated chemical potential $\mu_n$ to the Hamiltonian
\begin{equation}
H=H_1+H_J+\sum_n \mu_n (s_n^{\dagger}s_n+
\sum_{\alpha=x,y,z} t_{n, \alpha} t_{n, \alpha}^{\dagger}-1).
\label{h1J}
\end{equation}
The mean field Hamiltonian, $H_{mf}=\la H \ra $ is obtained by taking
$\la s_n \ra=\la s_n^{\dagger} \ra=\bar{s}$ and a global value for the
chemical potential, $\mu_n=\mu$.  After performing the Fourier
transform and the Bogoliubov transformation the mean field Hamiltonian
is written in the following form,
\be
H_{mf}=E_0+\sum_k \sum_{\alpha=x,y,z} 
\omega_{\alpha}(k) a_{k, \alpha}^{\dagger}  a_{k, \alpha}.
\label{hmf}
\ee
In this equation, $E_0$ is the ground state energy,
$\omega_{\alpha}(k)$ is the excitation energy of the new bosons
defined by the number operator $a_{k, \alpha}^{\dagger} a_{k,
\alpha}$. The new bosons can be expressed in terms of the bond
operators via the Bogoliubov transformations,
\bea
a_{k, \alpha}=\cosh(\theta_ {k, \alpha}) t_{k, \alpha} 
+\sinh(\theta_{k, \alpha}) t_{-k, \alpha}^{\dagger}, \nonumber \\
a_{-k, \alpha}^{\dagger}=\sinh(\theta_{k, \alpha})  t_{k, \alpha}
+\cosh(\theta_{k, \alpha}) t_{-k, \alpha}^{\dagger},
\label{transformation}
\eea
where $k$ is the momentum and 
\be
\theta_{k,x}=\theta_{k,y}=\frac{2 f_x(k)}{d_x(k)} ;\hspace{10mm}
\theta_{k,z}=\frac{2 f_z(k)}{d_z(k)}.
\ee
The functions $f_{\alpha}(k)$ and $d_{\alpha}(k)$ are defined in terms
of the coupling constants by
\bea
 f_x(k)= f_y(k)=\frac{t \bar{s}^2}{4} \gamma(k); 
\hspace{5mm}d_x(k)=d_y(k)=\mu+\frac{J_z}{4}+
\frac{t \bar{s}^2}{2} \gamma(k), \nonumber \\
f_z(k)=\frac{t \bar{s}^2}{4} \delta \gamma(k); 
\hspace{5mm}d_z(k)=\mu+\frac{2J_x-J_z}{4}+
\frac{t \bar{s}^2}{2} \delta \gamma(k),
\label{fd}
\eea
where $J_z=J \Delta$, $\gamma(k)=\sum_{i=1}^{D} \cos(k_i)$ and $D$ is the dimension of
rectangular Brillouin zone (BZ) associated with the D-dimensional
rectangular lattice. The bosonic excitation energies in
Eq.~(\ref{hmf}) are obtained as
\be
\omega_{\alpha}(k)=\sqrt{d_{\alpha}^2(k)-4f_{\alpha}^2(k)}, 
\hspace{5mm}\alpha=x, y, z.
\label{omega}
\ee
The ground state energy in $D$ dimension is then given by
\be
E_0=N\Big(\mu (\bar{s}^2-1)-(\frac{2J_x+J_z}{4})\bar{s}^2\Big)+
\frac{1}{2}\sum_k \sum_{\alpha=x,y,z}
\Big(\omega_{\alpha}(k)-d_{\alpha}(k)\Big).
\label{e0}
\ee
Minimization with respect to $\mu$ and $\bar{s}$ implies
\be
\frac{\partial E_0}{\partial \mu}=0, \hspace{5mm} 
\frac{\partial E_0}{\partial \bar{s}}=0.
\label{mfsce}
\ee
Explicitly this leads to 
\bea
\frac{1}{2N}\sum_{k\alpha}\frac{d_\alpha(k)}{\omega_{\alpha}(k)}
&=&(\frac{5}{2}-\bar{s}^2)\nonumber\\
\frac{t}{2N}\sum_{k\alpha}[\frac{d_\alpha(k)-2f_\alpha(k)}{\omega_{\alpha}(k)}
\gamma_\alpha(k)]&=&2J_x[\frac{1}{2}(1+\frac{J_z}{2J_x})-\frac{\mu}{J_x}].
\label{selfcons}
\eea
Here we defined $\gamma_{x,y}(k)=\gamma(k)$ and  $\gamma_z(k)=\delta\gamma(k)$. 
These equations will be solved selfconsistently for the average Kondo
singlet amplitude $\bar{s}$ and the chemical potential $\mu$.
It is worthwile to write the m.f. Hamiltonian Eq.(\ref{hmf}) in a
different form as 
\bea
H_{mf}&=&\tilde{E}_0+\sum_k \sum_{\alpha=x,y,z} 
\omega_{\alpha}(k) (a_{k, \alpha}^{\dagger} a_{k,
\alpha}+\frac{1}{2})\nonumber\\
\tilde{E}_0&=&-N\Big(\frac{2J_x+J_z}{4}(\bar{s}^2+\frac{1}{2})+\mu(\frac{5}{2}-\bar{s}^2)\Bigr)
\label{hmf2}
\eea
Because $\bar{s}\leq 1$ the first part ($\tilde{E}_0$) is always negative and
represents the m.f. singlet condensation energy. The second part is
the (positive) energy of excited triplet bosons. The ground state has only
a contribution from zero point motion. Both m.f. condensation energy
and energy of zero point quantum fluctuations contribute to the total
energy. Their balance determines the critical t$_c$ for the quantum
phase transition where the zero point fluctuation amplitude of bosons
diverges, i.e., $\omega_{\alpha}(Q)$ becomes soft. 
%
%
\begin{figure}
\begin{center}
\includegraphics[width=9cm,angle=-90]{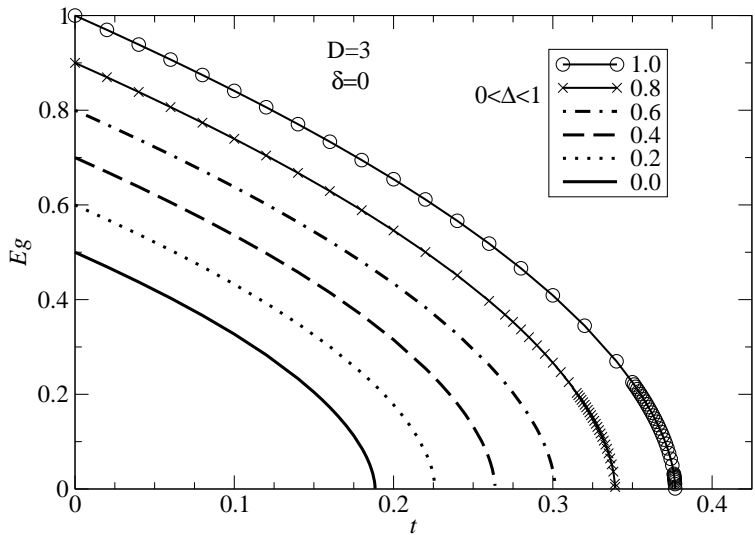}
\caption{The energy gap $\gap$ versus hopping strength in the three dimensional
lattice. The gap scales like $\gap\sim |t-t_c|^\nu$ with $\nu\simeq
0.55$.}  
\label{fig4}
\end{center}
\end{figure}
%

\section{Numerical results: The quantum critical phase diagrams}
\label{sec4}

We will now obtain the numerical solutions to the coupled equations
(\ref{mfsce}). For this purpose we first consider the XY- case, i.e.,
$\delta=0$ and arbitrary anisotropy $\Delta$ for the Kondo term in
Eq.~(\ref{hamiltonian}). We remind that in D=1 the XY-case is
equivalent to the genuine Kondo lattice model with 
free itinerant electrons
of a band-width W=4t. Later we will also investigate the case of 
nonzero $\delta$.
%
\begin{figure}
\begin{center}
\includegraphics[width=9cm,angle=-90]{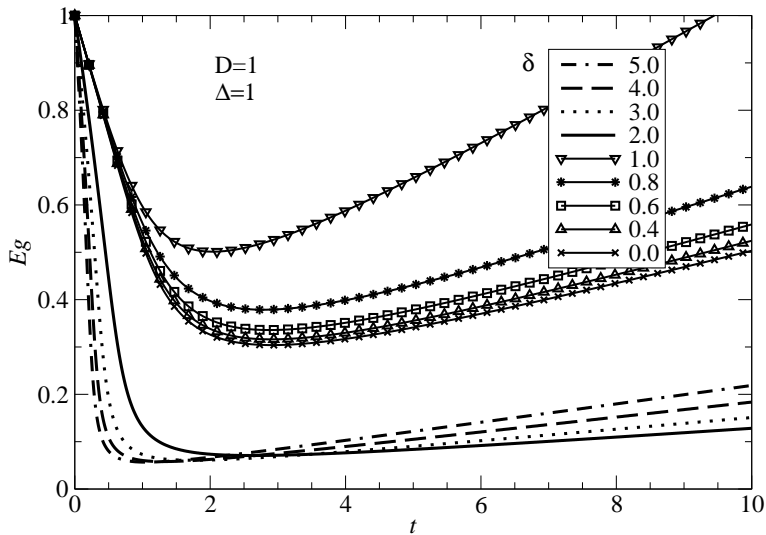}
\caption{The energy gap versus $t$ for different values of $\delta$ in the one
dimensional lattice. The gap is always nonzero for $ \delta \ge 0$.
}
\label{fig5}
\end{center}
\end{figure}
%
\subsection{XY-case: $\delta=0$}

In this case the z-polarised branch of excitations has a
dispersionless value $\omega_z(k)=\omega_0$. However, the other two
branches of excitations ($\omega_x=\omega_y$) show a dispersion which
has a minimum at the antiferromagnetic reciprocal vector at the corner
of the BZ, e.g., Q=($\pi$,$\pi$) for D=2. The
minimum value of excitations defines the energy gap
\be
\gap=(\mu+\frac{J_z}{4})\sqrt{1-D(\frac{t \bar{s}^2}{\mu+J_z/4})}.
\label{gap}
\ee
In the present case the energy gap depends on the two parameters $t$
and $J_z$. The gap $\gap$ defines the energy scale for the Kondo
singlet phase.  When this energy scale approaches zero the other type
of excitations become important and the model will encounter a phase
transition from Kondo singlet to the antiferromagnetic phase.  We will
see that there exists a critical value for $t=t_c$ where this phase
transition occurs in $D=2,3$ while no transition happens in
$D=1$. This is similar to the results in Ref. \onlinecite{Zhang00} for
the special case ($\delta$,$\Delta$)=(0,1) and to the previous
results of MC simulations \cite{Scalettar85} for the 1D model. 
We have solved the mean field equations (\ref{mfsce}) numerically for
general values of the coupling constants ($t, J$) and
anisotropies ($\delta$, $\Delta$). The resulting value of $\mu$ and
$\bar{s}$ are replaced in Eq.(\ref{gap}) to obtain the energy gap and
track the location $t_c$ where it vanishes.\\
%
\begin{figure}
\begin{center}
\includegraphics[width=9cm,angle=-90]{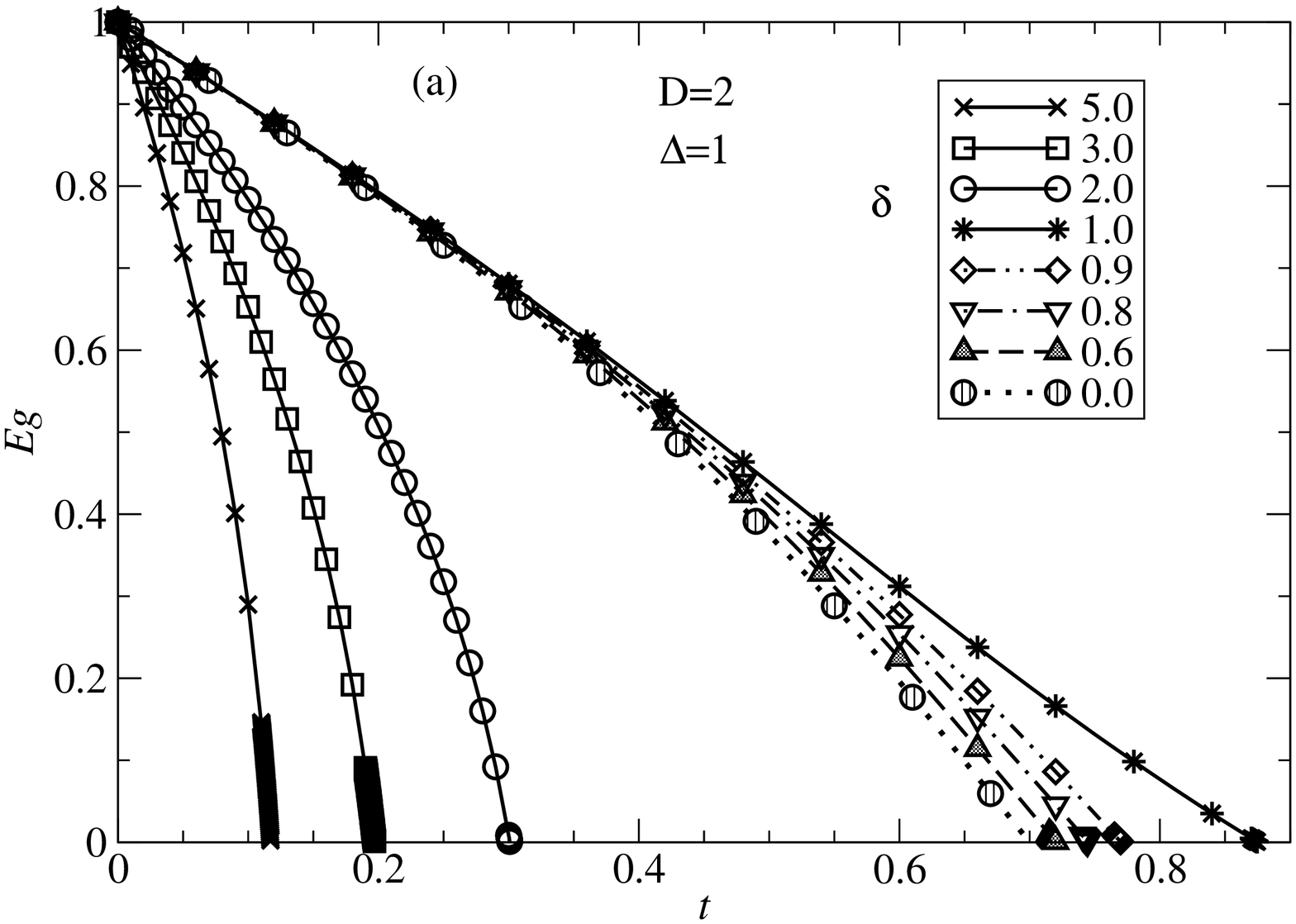}
\includegraphics[width=9cm,angle=-90]{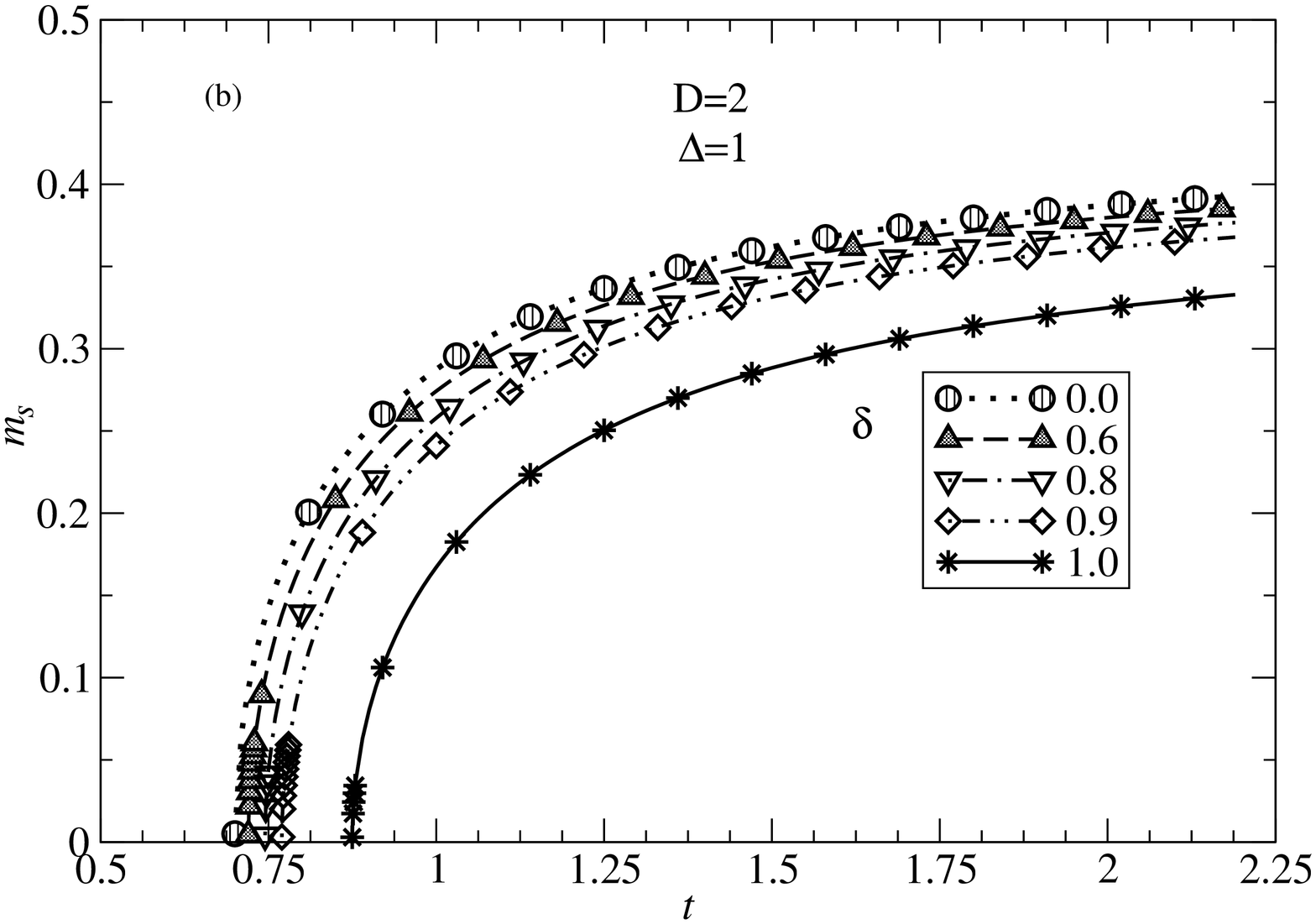}
\caption{
Two dimensional lattice: (a) The behavior of gap for two different
regimes of $\delta$ versus $t$ in two dimensional lattice for
$\Delta=1$. The scaling of gap close to the critical hopping ($t_c$)
is obviously different for $0 \le \delta \le 1$ and $\delta > 1,
(\delta=2, 3, 5)$.  (b) Staggered magnetization ($m_s$) versus $t$ in
the antiferromagnetic phase, just above the quantum critical point at
t$_c$.  The local anisotropy is $\Delta=1$ and the itinerant one
varies within $0 \le \delta \le 1$.}
\label{fig6}
\end{center}
\end{figure}
%
In Fig.\ref{fig1} we have plotted the energy gap versus $t$ (the hopping strength)
for different anisotropies $0 \leq \Delta=\frac{J_z}{J_x} \leq 1$ in $D=1$. 
The value of gap is reduced
by decreasing the anisotropy from the isotropic ($\Delta=1$) case to
the XY-type ($\Delta=0$) Kondo interaction.  However, it is always
nonzero and never vanishes. Thus, the effect of anisotropy does not
change the universality behavior of the $D=1$ case where the system is
always in the Kondo singlet phase. We have also considered values with
$\Delta > 1$ (easy axis- or Ising-type regime) and found a similar
behavior as in Fig.\ref{fig1}, provided we change the energy scale
from $J_x$ to $J_z$. It can be seen from the expression for the energy
gap that there is a duality between the 
$0\leq \Delta \leq 1$ and $\Delta > 1$ regions by interchanging $J_x
\longleftrightarrow J_z$.  For theoretical reasons we might also
consider the negative values of anisotropy ($-1 \le \Delta \le 0$).
Its behavior is  similar to Fig.\ref{fig1} while the minimum of
gap is reduced and vanishes at $t=0$ for $\Delta=-1$. Actually, at
$\Delta=-1$ the energy difference of the local singlet and the two
triplet states becomes zero which leads to the vanishing of the energy
gap. However, in this limit the above mean field approach is not reliable 
where the triplet contribution can not be neglected in the ground state.
For negative values of $\Delta$ the
average value of the triplets is comparable with the singlets and
should be taken into account on the same footing.\\
%
\begin{figure}
\begin{center}
\includegraphics[width=9cm,angle=-90]{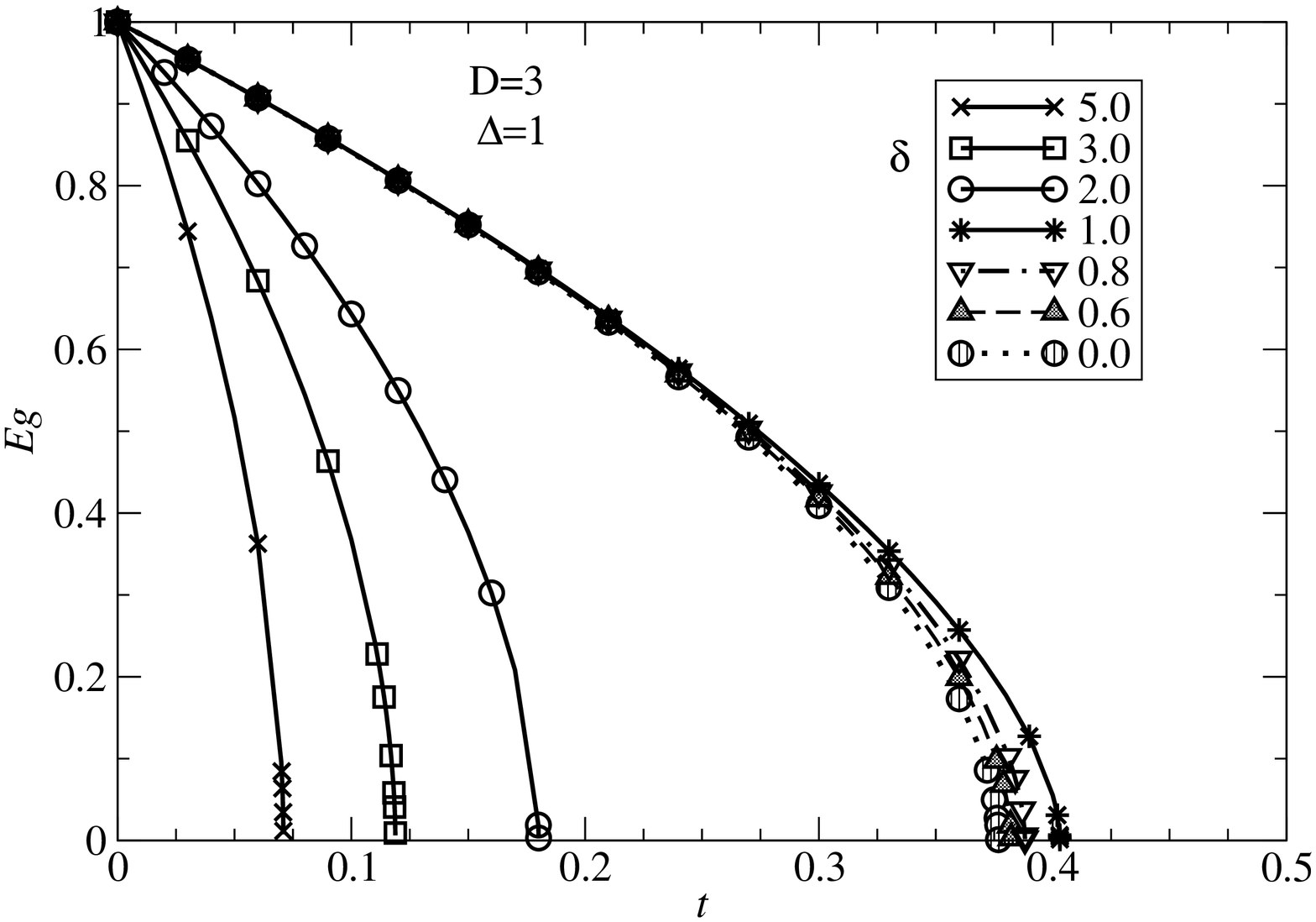}
\caption{The effect of $\delta$ on the behavior of gap versus $t$ for the three
dimensional cubic lattice.  The gap vanishes at the critical point
with different scaling behavior for $0 \le \delta \le 1$ and $\delta > 1,
(\delta=2, 3, 5)$.  }
\label{fig7}
\end{center}
\end{figure}
%
Let us now present the results for the square lattice ($D=2$) shown in
Fig.\ref{fig2}.  We have plotted the energy gap versus $t$ for
different values of $0 \le \Delta \le 1$.  For all values of $\Delta$
the gap vanishes at the critical point $t_c$, where the transition
from Kondo singlet to the antiferromagnetic phase occurs. The
dependence of the critical point on the anisotropy is linear and is
fitted well by $t_c \simeq 0.35 J (\Delta+1)$ which is also
valid for negative values of $-1 \le \Delta \le 0$ but has not been
shown here.  We will derive an equation for the critical temperature of $D=2, 3$
in terms of the coupling parameters 
in Sec. \ref{af-phase}. The linear dependence of $t_c$ on $\Delta$ shows that the
anisotropy term is important to stabilize the Kondo singlets. Thus,
for higher $\Delta$, higher value of the hopping term is required to
destroy the Kondo singlet and induce the antiferromagnetic order.
However, the qualitative behavior of the energy gap  versus $t$ is the same
for all $\Delta$. Our fine tuned data plotted in the inset of  Fig.\ref{fig2}  show 
that the gap vanishes close
to $t_c$ with a critical exponent $\nu$ according to
\begin{equation}
\gap \sim |t-t_c|^{\nu},
\end{equation}
where $\nu\simeq 1$ for $D=2$. The exponent is the same for all
anisotropies and this is in agreement with the observation that the
local interaction terms will not change the universal behavior.  
The log-log plot of $E_g$ versus $|t-t_c|$ for all $\Delta$ fall on each
other which justifies the scaling behaviour close to critical point.
The
extension of our calculation to $\Delta >1$ (Ising regime) shows
qualitatively similar results.  We have also plotted the value of
$\bar{s}=<s>$ versus the hopping term in Fig.~\ref{fig3} for two
different values of anisotropy, $\Delta=0, 1$ in the case of two
dimensional square lattice. These data justify our assumption that
the expectation value of singlets in the ground state is close to one
up to the critical $t_c$ value. It supports our claim to neglect the
occupation of triplets in the ground state in the mean field approach.\\

Our results for $D=3$ are shown in Fig.~\ref{fig4}. The gap versus $t$
shows similar behavior qualitatively for all values of $\Delta$. There
exists always a critical point ($t_c$) which shows the transition from
Kondo singlet to the antiferromagnetic phase. The critical point depends
on the anisotropy linearly, $t_c \simeq 0.188 J (\Delta+1)$. The exact expression
for the critical point is given by an equation in Sec.\ref{af-phase}. However, the
dependence is weaker than in the case of $D=2$. The reason is related to
the increase of the number of nearest neighbor sites which allows
hopping more easily. Thus, the hopping term is more effective here.  The analysis
of log-log plot of Fig.~\ref{fig4} close to the critical point gives
the gap exponent to be $\nu\simeq 0.55$ for the three dimensional
cubic lattice.

\subsection{general ansisotropic case $\delta\neq0$}

Our results in the previous case, $\delta=0$, show that the anisotropy
in the local interaction ($\Delta$) does not change the qualitative
behavior of the necklace model on the cubic lattices with D=1-3. In
order to study the effect of a nonzero anisotropy $\delta\neq0$ in the
'itinerant' part ($\sim t$) in Eq.~(\ref{hamiltonian}) it is therefore
sufficient to consider only the isotropic case of the local Kondo
interactions, i.e., $J_x=J_z=J$, ($\Delta=1$).

When $\delta$ is nonzero, we find three excitation modes in
Eq.(\ref{omega}) which now all show dispersion.  The minimum of
excitations which introduces the energy gap appears again at the
antiferromagnetic reciprocal wave vector Q.  In contrast to the previous
case, the excitation in z-direction ($\omega_z$) is important and
defines the gap for the easy-axis type anisotropy $\delta > 1$. The minimum
excitation energies for the modes at Q  are given by
\bea
\gapx=\gapy=(\mu+\frac{J}{4})\sqrt{1-D(\frac{t \bar{s}^2}{\mu+J/4})},\no \\
\gapz=(\mu+\frac{J}{4})\sqrt{1-\delta D(\frac{t \bar{s}^2}{\mu+J/4})}.
\label{gapzx}
\eea
For $0 \le \delta < 1$, we have $\gapx < \gapz$. However, at
$\delta=1$ the energy gap of all excitations coincide. If we consider
$\delta > 1$ the situation is reversed and $\gapz < \gapx$.  This is
important because the vanishing of gap and the appearance of soft modes
define the transition from Kondo singlet to the antiferromagnetic phase.\\

In the case of one dimensional space ($D=1$) the energy gap is always
nonzero for $0 \le \delta \le 1$.  The dependence of the gap on the
anisotropy of the 'itinerant' term ($\delta$) in
Eq.(\ref{hamiltonian}) is smooth as shown in in Fig.\ref{fig5}. 
However, there is an abrupt change in the value of gap at $\delta=1$ which 
is related to the contribution of the third soft mode ($\omega_z$) to the lowest 
energy part of the model.
The situation is almost the same for $\delta > 1$ where the gap never
vanishes and therefore no phase transition appears for increasing $t$.

The effect of $\delta$ anisotropy in the two dimensional lattice is
plotted in Fig.\ref{fig6}(a).  As discussed earlier the gap $\gap$ is
defined by the $\gapx$ and $\gapy$ for $0 \le \delta \le 1$ and
$\gapz$ for $\delta > 1, (\delta=2, 3, 5)$. For $\delta <1$ the effect
of anisotropy is weak and it changes the critical point ($t_c$)
slightly. The dependence of the critical point ($t_c$) is not a simple
function as will be given in the next section.
The qualitative behavior is the same for all $\delta <1$ and the gap
exponent is approximately $\nu \simeq 1$. We observe a jump in the
critical point which appears at $\delta=1$.  Because at this point
another soft mode ($\omega_z$) is added to the excitations at the
critical point which needs a larger hopping strength $t$ to overcome
them. If we describe the gap close to the critical point by
$\gap=A(\delta) |t-t_c|^\nu$ the coefficient $A(\delta=1)$ is almost
half the value for the other cases ($A(\delta\neq1)$). When $\delta
>1$, $t_c$ is reduced and the Kondo singlet phase is
limited to a smaller region. 
In this case the interaction between the itinerant electron ($t \delta$) 
dominates the Kondo term ($J$) for smaller values of $t$.
The change of universality in
approaching the critical point is also obvious from Fig.\ref{fig6}(a).

Increasing $t$ beyond $t_c$ leads to antiferromagnetic order as
discussed in Sect.~\ref{af-phase}. For comparison we have also plotted
the AF order parameter (staggered magnetization) versus $t$ in
Fig.\ref{fig6}(b). The staggered magnetization is zero in the singlet
phase until $t=t_c$. For high values of $t$ ($t \gg J$), the
antiferromagnetic order parameter is close to saturation. The
reduction of $t$ increases the effect of the local exchange
interaction which favors the singlet phase.  At the critical point
($t=t_c$) the staggered magnetization is destroyed by the AF
fluctuations connected with the soft modes. This happens exactly at
the point where the Kondo singlet energy gap mentioned in previous
paragraph vanishes.

We also show the effect of nonzero $\delta$ on the three
dimensional cubic lattice in Fig.\ref{fig7}.  As discussed earlier,
the minimum of excitations in x- or y-direction gives the scale of
energy ($\gapx,\gapy$) for $0 \le \delta \le 1$ and $\gapz$ for
$\delta >1$. The effect of $\delta$ on the behavior of energy gap
versus $t$ is weaker than the case of $D=2$. The gap vanishes at the
critical point which has a slight dependence on $\delta$. There is a
jump on the critical point at $\delta=1$ which is the result of three
soft modes at this point while for $0 \le \delta \le 1$ there are two
soft modes and for $\delta >1$ only one soft mode.  Although $\delta$
is a global term in the interaction there is no change in the scaling
behavior of the gap close to the critical point. The gap exponent for
$0 < \delta \le 1$ is the same as for zero $\delta$, namely $\nu\simeq
0.55$. However, we have observed different behavior for $\delta >1$
where the gap vanishes close to critical point more rapidly. But we were
not able to find a scaling exponent in this regime because here $t_c$
varies too strongly with $\delta$. In the other cases we have determined the
scaling exponents numerically from the selfconsistent solutions and
found them to lie close to $\nu=\frac{1}{2}$ for 2D and $\nu =1$ for
3D. The deviations are mostly due to the numerical inaccuracy of $t_c$
determination. This is the expected result for the present mean field
approach. In the special case ($\delta,\Delta$)=(0,1) we have expanded 
the selfconsistent equations Eq.(\ref{selfcons}) around $E_g$=0 which indeed leads
to the analytical mean field scaling behaviour for $E_g$.

\section{Antiferromagnetic phase}
\label{af-phase}
\begin{figure}
\begin{center}
\includegraphics[width=9cm,angle=0]{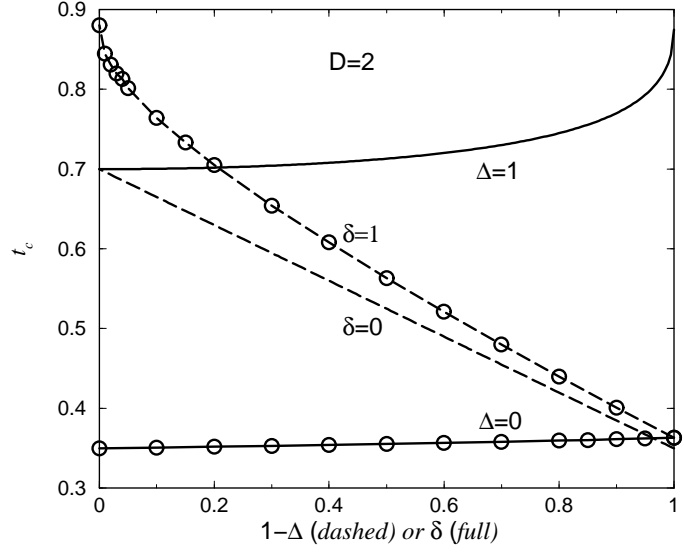}
\caption{Dependence of the quantum critical point $t_c$ (in units of J$_x$) for $D=2$
on itinerant ($\delta$, full line) and local (1-$\Delta$, dashed line)
anisotropy parameters. The lines without symbols are from the closed
analytical form in Eq.(\ref{tcexpl}), the curves with symbols are
obtained from numerical solution Eq.(\ref{mustbar}) or Eq.(\ref{selfcons}).}
\label{fig8}
\end{center}
\end{figure}

The mean field approach can be simply extended to the antiferromagnetic phase
for $D=2,3$. For large enough $t/J$ the long range antiferromagnetic order sets up
which means the condensation of a component of local spin triplets. We assume the 
condensation in the $x$-component of the spin triplet ($t_{k,x}$) at the antiferromagnetic
wave vector ($Q=(\pi,\pi)$ for D=2),
\begin{equation}
t _{k, x}=\sqrt{N}\bar{t}\delta_{k, Q} + \eta_{k, x}.
\label{tq}
\end{equation}
In Eq.(\ref{tq}) $\bar{t}$ is the mean value of the $x$-component of the spin triplet in
the ground state and $\eta_{k, x}$ is its quantum fluctuations. We replace $t_{k, x}$ into the
Fourier trasformed of Eq.(\ref{h1J}) to derive the mean field Hamiltonian assuming 
$\la s_k \ra=\bar{s}$ and $\la t_{k, x} \ra=\bar{t}$. The mean field Hamiltonian will 
be in the following form,
\begin{equation}
H^{AF}=E_0^{AF}+\sum_k \omega_z(k) t_{k,z}^{\dagger}t_{k,z}
+\sum_k \omega_x(k) (\tilde{\eta_k}^{\dagger}\tilde{\eta_k}+ \tilde{t}_{k,y}^{\dagger}\tilde{t}_{k,y}),
\end{equation}
where $\tilde{O}$ represent the new operators after Bogoliubov transformation.
The ground state energy ($E_0^{AF}$) is defined below,
\begin{equation}
E^{AF}_0=E_0+N \bar{t}^2(\mu+\frac{J_z}{4}-D t \bar{s}^2),
\end{equation}
where $E_0$ has been defined in Eq.(\ref{e0}). The ground state energy $E^{AF}_0$  is minimized
with respect to $\mu$, $\bar{s}$ and $\bar{t}$ which leads to the following equations,
\begin{eqnarray}
\mu&=&D t \bar{s}^2-\frac{J_z}{4},\nonumber\\
\bar{s}^2&=&\frac{5}{4}+\frac{J_x+J_z}{4 D t}-\frac{1}{2 N}\sum_k \sqrt{1+\frac{\gamma(k)}{D}}
-\frac{1}{4 N}\sum_k \frac{X_{+}^{k}}{\omega_z(k)}\nonumber\\
\bar{t}^2&=&\frac{5}{4}-\frac{J_x+J_z}{4 D t}-\frac{1}{2 N}\sum_k \frac{1}{\sqrt{1+\frac{\gamma(k)}{D}}}
-\frac{1}{4 N}\sum_k \frac{X_{-}^{k}}{\omega_z(k)}.
\label{mustbar}
\end{eqnarray}
Furthermore,
\begin{eqnarray}
X_{+}^{k}&=&\frac{1}{2}(J_x-J_z)(1+\delta \frac{\gamma(k)}{2 D})+ D t \bar{s}^2(1+\delta \frac{\gamma(k)}{D}),\nonumber\\
X_{-}^{k}&=&\frac{1}{2}(J_x-J_z)(1-\delta \frac{\gamma(k)}{2 D})+ D t \bar{s}^2.
\label{xpm}
\end{eqnarray}
Generally the equation for $\bar{s}$ should be solved selfconsistently
and then $\mu$ and $\bar{t}$ can be found directly from the above
equations.  However, for $\Delta=1$, the above equation
(Eq.(\ref{mustbar})) for $\bar{s}$ is simplified to a direct integral
expression. If in addition $\delta =0$ the result in
Ref.\onlinecite{Zhang00} is recovered.  For the bipartite lattice
which we have assumed the antiferromagnetic order parameter is defined
on A- or B-sublattices as,
\begin{equation}
\la S_x \ra_A= - \la S_x \ra_B= \bar{s} \bar{t},
\end{equation}
Here $m_s$=$\bar{s}\bar{t}$ is the staggered moment. It is plotted for
$D=2$ and different $\delta$ values in Fig.~\ref{fig6}(b). The
appearance of long range antiferromagnetic order (nonzero $m_s$)
defines the quantum critical point $t=t_c$. This is identical to the
value where the Kondo singlet gap vanishes as seen in the conjugate
plot in Fig.~\ref{fig6}(a). The formulation presented in this section
allows us to obtain an explicit expression for the critical point
($t_c$) for the special cases ($\delta,\Delta$)= ($0,\Delta$) and
($\delta,\Delta$)= ($\delta,1$)which is given by
\begin{equation}
\frac{1}{t_c}=(\frac{2 D}{1+\Delta})
\Big( \frac{5}{2}-\frac{1}{2 N}\sum_{k,
\alpha}\frac{1}{\sqrt{1+\frac{\gamma_{\alpha}(k)}{D}}} \Big).
\label{tcexpl}
\end{equation}
Here we used again $\gamma_{x,y}(k)=\gamma(k)$ and
$\gamma_z(k)=\delta\gamma(k)$. For the other cases one has to obtain
$t_c$ numerically by the selfconsistent solution of Eq.(\ref{mustbar})
or equivalently by Eq.(\ref{selfcons}).  For D=2 we have plotted the
value of $t_c$ versus $1-\Delta$ (at fixed $\delta = 0$) or versus
$\delta$ (at fixed $\Delta = 1$) in Fig.~\ref{fig8}. The bare lines
correspond to the analytical solution of Eq.(\ref{tcexpl}) while the
lines with symbols correspond to the numerical solution for $t_c$. One
notices that on approaching ($\delta,\Delta$) $\rightarrow$ (1,1)
where $t_c$=0.88 (the two uppermost points in Fig.~\ref{fig8}) the
slope of $t_c$($\Delta$) or t$_c$($\delta$) diverges. This signifies
the change of universality class from U(1) to SU(2) at the point
($\delta,\Delta$)=(1,1) which leads to the apperance of a third soft
mode at $t_c$.
%
%
\section{Summary and discussions}
\label{sec5}

We have studied the quantum critical properties of the fully
anisotropic Kondo-necklace model in dimension D=1-3. We used the
bond-operator representation of spin variables with constraints
implemented in the mean field approximation. For general anisotropy
ratios ($\delta$, $\Delta$) we have calculated the critical values
$(t/J)_c$ and the scaling exponenets of the Kondo singlet gap close to
the quantum critical point which correspond to the mean field
exponents. 

We find that in general the anisotropy $\Delta$ of the
local Kondo term influences the qualitative behavior of the singlet
gap only little, although the value of the critical ratio $(t/J)_c$
depends strongly on $\Delta$. If $\Delta$ decreases the Kondo singlet
is destabilized and the critical value $(t/J)_c$ decreases.  On the
other hand the anisotropy of the inter-site term which mimics the
itinerant electrons has a weak effect on the position of the critical
point (Fig.~\ref{fig8}) and leaves the universal behavior similar as
in the conventional XY-type ($\delta$ = 0) Kondo necklace model. On
approaching the SU(2) Heisenberg case ($\delta$, $\Delta$)=(1,1) $t_c$
exhibits a singular behaviour as function of anisotropies.  This is
most pronounced in D=2 where also the scaling coefficient (not the exponent) of
the gap close to the QCP changes considerably as compared to the $\delta$ = 0
case. Furthermore the value of the critical $(t/J)_c$ with properly
defined scale $J$ exhibits a jump at $\delta$=1. This peculiar effect
is due to a mode crossing of the x,y and z branches of excitations at
the AF wave vector as function of $\delta$. 

We have also derived and
solved the selfconsistency equations on the magnetic side for general
($\delta$, $\Delta$). This allows us to give an explicit expression for
the quantum critical point value $t_c$ as function of anisotropy
parameters ($\delta$, $\Delta$) for the cases $\delta$=0 or $\Delta$=1
for dimension D=2,3. To discuss the scaling exponents beyond mean field
approximation the renormalisation of triplet excitation energies close
to the QCP caused by fluctuations in the bosonic variables has to be
included.

\section{Acknowledgment}
The authors would like to thank Burkhard Schmidt and Ivica Zerec for
discussions.

\section{Appendix}

The explicit form of $H_2$ and $H_3$ introduced in Sect.~\ref{sec3} in terms of bond operators
are as follows:
\bea
H_2&=&\frac{-t}{4}\sum_{\la n,m\ra}\Big((t^{\dagger}_{n,y} t_{n,z}- h.c.)
(t^{\dagger}_{m,y} t_{m,z}- h.c.)
+(t^{\dagger}_{n,x} t_{n,z}- h.c.)(t^{\dagger}_{m,x} t_{m,z}- h.c.) \nonumber \\
&+&\delta(t^{\dagger}_{n,x} t_{n,y}- h.c.)
(t^{\dagger}_{m,x} t_{m,y}- h.c.)\Big).
\eea
The terms appearing in $H_2$ describe the interaction between triplets.
Since the triplet occupation is very small in the ground state, the effect 
of $H_2$ on the mean field result is negligible.

The remaining part of $H_t$ is expressed by $H_3$ which gives zero contribution to 
the ground state energy in the mean field approximation.
\bea
H_3&=&\frac{i t}{4}\sum_{\la n,m\ra}\Big([
(s_n^{\dagger}t_{n,x}+t_{n,x}^{\dagger}s_{n})(t^{\dagger}_{m,y} t_{m,z}-t^{\dagger}_{m,z} t_{m,y})
+(s_n^{\dagger}t_{n,y}+t_{n,y}^{\dagger}s_{n})(t^{\dagger}_{m,z} t_{m,x}-t^{\dagger}_{m,x} t_{m,z})
\nonumber \\
&+&\delta(s_n^{\dagger}t_{n,z}+t_{n,z}^{\dagger}s_{n})(t^{\dagger}_{m,x} t_{m,y}-t^{\dagger}_{m,y} t_{m,x})]
+ h.c.\Big).
\eea



\end{document}